\documentclass[sigconf]{acmart}
\usepackage{subcaption}
\usepackage{balance}

\usepackage{algorithm}
\usepackage{algpseudocode}
\usepackage{amsmath}

\AtBeginDocument{%
  }

\setcopyright{acmlicensed}
\copyrightyear{2025}
\acmYear{2025}
\acmDOI{XXXXXXX.XXXXXXX}
\acmConference[SIGSPATIAL '25]{Minneapolis, USA}
\acmISBN{978-1-4503-XXXX-X/2025/06}

\begin{document}

\title{Spatial Regionalization: A Hybrid Quantum Computing Approach}

 \author{Yunhan Chang$^{1,2}$}
 \affiliation{%
   \department{$^1$Department of Computer Science and Engineering}
   \position{$^2$Center for Geospatial Sciences}
   \institution{University of California, Riverside}
    \country{Riverside, CA, USA}
 }
 \email{ychan268@ucr.edu}

 \author{Amr Magdy$^{1,2}$}
 \affiliation{%
   \department{$^1$Department of Computer Science and Engineering}
   \position{$^2$Center for Geospatial Sciences}
   \institution{University of California, Riverside} 
    \country{Riverside, CA, USA}
 }
 \email{amr@cs.ucr.edu}

 \author{Federico M. Spedalieri$^{3}$}
 \affiliation{
  \position{$^3$Information Sciences Institute}
  \institution{University of Southern California}
   \country{Los Angeles, CA, USA}
 }
 \email{fspedali@isi.edu}

 \author{Ibrahim Sabek$^{4,5}$}
 \affiliation{
  \department{$^4$Department of Computer Science}
  \position{$^5$Spatial Sciences Institute}
  \institution{University of Southern California}
  \country{Los Angeles, CA, USA}
 }
 \email{sabek@usc.edu}


\begin{abstract}
Quantum computing has shown significant potential to address complex optimization problems; however, its application remains confined to specific problems at limited scales. Spatial regionalization remains largely unexplored in quantum computing due to its complexity and large number of variables.
In this paper, we introduce the first hybrid quantum-classical method to spatial regionalization by decomposing the problem into manageable subproblems, leveraging the strengths of both classical and quantum computation.
This study establishes a foundational framework for effectively integrating quantum computing methods into realistic and complex spatial optimization tasks.
Our initial results show a promising quantum performance advantage for a broad range of spatial regionalization problems and their variants.

\end{abstract}





\maketitle


\section{Introduction}

Quantum annealing~\cite{kadowaki1998}
is an optimization technique inspired by quantum mechanics, designed to solve complex optimization problems by exploring a vast solution space through quantum fluctuations.
Unlike its classical counterpart, simulated annealing~\cite{Koshka2020DWave} that is based on thermal fluctuations, quantum annealing uses quantum tunneling to escape local minima and potentially converge on the global optimum. 
In recent years, the quantum annealing hardware has been significantly advanced to support thousands of qubits, e.g., \textit{D-Wave Quantum Inc.} annealer currently offers approximately \textbf{5000~qubits}~\cite{King2023Quantum}.
However, current quantum systems remain constrained by notable technical limitations that limit solving problems of large sizes.
To address the scale challenge, quantum annealers are supplemented by hybrid solvers that integrate classical CPUs with Quantum Processing Units (QPUs) at different phases.
The classical computer handles problem decomposition, setting initial conditions, and refining results using well-established optimization methods, while the QPU leverages quantum annealing with its native QUBO-like formulations~\cite{Glover2022QUBO}. These solvers, such as DQM, CQM, and NL solvers~\cite{dwave_leap_hybrid}, can partition large-scale problems into smaller sub-problems that are then solved with the QPU.
Such advances have enabled quantum computers to handle complex optimization problems and enterprise-level applications~\cite{web:dwavesuccessstories}.

Spatial regionalization is an essential and expensive spatial operation that serves many real-world applications in economics, urban planning, and other domains requiring efficient spatial resource allocation~\cite{Alrashid2022SMP,Liu2022PRUC,Kang2022EMP}.
Regionalization aggregates spatial areas into contiguous regions, aiming to minimize heterogeneity within each region.
Constrained by explicit spatial contiguity requirements, spatial regionalization is proved to be NP-hard~\cite{Kang2022EMP}, and finding optimal solutions remains computationally infeasible due to combinatorial complexity~\cite{AlrashidLM23, Liu2022PRUC, Kang2022EMP}.

This paper introduces the first hybrid quantum-classical approach designed for seeding-based spatial regionalization.
We decompose regionalization into smaller, quantum-processable subproblems and explore the solution space more effectively than the traditional heuristic methods.
Specifically, we decompose seeding-based regionalization into two phases, spanning five stages, adapting a variation of our framework proposed in~\cite{Alrashid023} for classical computers.
Then, the quantum annealing is leveraged to improve the performance of two stages: \textbf{\textit{seed selection}} and \textbf{\textit{local optimization}}, which are used by all seeding-based regionalization techniques.
Therefore, our proposed improvements are general and applicable in various applications.
Our initial empirical results on the D-Wave quantum annealer show promising improvements in seed selection quality and local optimization runtime.
Ongoing and future explorations might exploit our quantum model in~\cite{amrQdata2025Paper} for enforcing spatial contiguity constraints, which are imposed by all regionalization techniques, to build a quantum-optimized \textit{region growing} stage.
The rest of this paper introduces a brief background, discusses the related work, and presents our proposed approach and initial empirical results.



\section{Background}


\textbf{Quantum annealing (QA)} has been proposed as a novel heuristic to solve many of the well-known NP-hard problems~\cite{kadowaki1998}, and it is similar to Simulated Annealing (SA). In SA, a solution space is explored by randomly proposing changes to the current solution. 
If the new value of the objective function is decreased, the proposed change is accepted.
If the objective function increases, the change is accepted with some probability according to the temperature.
In QA, these changes are applied on a quantum superposition of the different solutions, allowing for a massively parallel search of the solution space. This quantum parallelism, combined with the quantum tunneling~\cite{amrQdata2025Paper}, facilitates fast exploration and potentially avoids being stuck in local minima.

\noindent\textbf{\underline{Q}uadratic \underline{U}nconstrained \underline{B}inary 
\underline{O}ptimization (QUBO)}.
In QA, the problem is represented as a \textbf{QUBO formulation} that consists of a quadratic objective function over binary variables, while any constraints are represented as penalty terms~\cite{Glover2022QUBO}.
QUBO is an optimization technique that is inherently similar to other operations research techniques and requires no or limited background knowledge of the internals of the quantum annealing process from a user perspective. 
Under the hood, the quantum annealer encodes the QUBO formulation in the interactions between 
\textit{qubits}, which are the quantum counterpart of the classical binary bits.
More details about the quantum annealing process can be found in~\cite{amrQdata2025Paper}.

\noindent\textbf{Hybrid quantum-classical computing.}
Although many optimization problems can be formulated as QUBOs, the limitations in the current quantum hardware, e.g., small size and low degree of qubit connectivity, restrict the size and complexity of problems that can be practically solved.
This challenge is further amplified in problems with many constraints, where encoding constraints as penalty terms~\cite{Glover2022QUBO} often requires additional slack variables, consuming even more qubits.
A widely adopted and practical approach is integrating quantum annealers within a hybrid framework, combining both classical CPU and quantum computing powers to solve large problems efficiently using \textbf{\textit{hybrid solvers}}. 
The core idea of a \textit{hybrid solver} is to decompose large or dense problems into smaller subproblems that fit within the connectivity and size constraints of the quantum hardware. In this framework, a quantum processor is paired with a classical heuristic in an iterative loop: the quantum processor solves selected subproblems, while the classical component manages orchestration, constraint handling, and post-processing. By offloading constraint enforcement and much of the solution space exploration to the classical side, hybrid solvers effectively mitigate the challenges posed by sparse qubit connectivity.
\textit{D-Wave Quantum Inc.}, a leading provider of quantum annealing hardware, currently offers four hybrid solvers tailored to different problem types and variable domains: the Binary Quadratic Model (BQM), the Discrete Quadratic Model (DQM), the Constrained Quadratic Model (CQM), and the Nonlinear-Program Hybrid Solver (NL-solver)~\cite{dwave_leap_hybrid}.
Notably, both CQM and NL-solver support the explicit expression of constraints in a declarative form.

\vspace{-8pt}
\section{Related Work} 
\label{sec:relatedwork}

\textbf{Classical spatial regionalization} clusters a collection of spatial areas, i.e., polygons, into regions that are spatially contiguous and homogeneous with respect to one or more attributes~\cite{AlrashidLM23,Alrashid2022SMP,Liu2022PRUC,Kang2022EMP,Alrashid023, web:geoda_spatial_clustering}.
Numerous variants exist, for example, the p-regions problem takes the number of regions \textit{p} as user input, the max-p-regions problem automatically finds the maximum number of regions, whereas the p-compact-regions problem emphasizes spatial compactness of regions.
All regionalization variants are NP-hard, so practical solutions rely on heuristics to find approximate \textit{good} solutions for large data.
The two major classes of classical solutions are:
(a)~\textbf{seeding-based regionalization} and 
(b)~\textbf{graph-based regionalization}. 
Seeding-based regionalization techniques select seed areas and build regions, i.e., clusters, around these seeds~\cite{Alrashid023,web:geoda_spatial_clustering}.
Graph-based regionalization builds connected regions by working directly on spatial adjacency graphs~\cite{web:geoda_spatial_clustering}.
It includes both top-down, e.g., SKATER, and bottom-up agglomerative techniques, e.g., REDCAP~\cite{web:geoda_spatial_clustering}.
Over the past decade, it has been empirically shown that seeding-based techniques have better scalability than graph-based techniques, so they empower spatial data analysts to support large datasets~\cite{Liu2022PRUC,Kang2022EMP,AlrashidLM23,Alrashid2022SMP}.
However, classical heuristics are computationally extensive and prone to getting trapped in local optima, making the search for high-quality solutions expensive.
By using quantum tunneling, quantum annealing can achieve high-quality solutions faster.
None of the existing techniques leverages quantum annealing to address this NP-hard problem.

\noindent\textbf{Quantum-based spatial clustering} has emerged as a promising tool for NP-hard clustering and partitioning problems.
Mapping spatial clustering to QUBO form is non-trivial, yet several studies show feasibility for related graph-partitioning tasks.
Ushijima-Mwesigwa~\cite{QAGraph2017Ushijima} encodes modularity-maximizing community detection as a QUBO, with each node’s binary spin indicating community membership.
Negre~\cite{QACommunity2020Negre} extends this to $k$ communities on a D-Wave 2000Q, obtaining modularity scores comparable to leading classical algorithms.
Arthur and Date~\cite{Quantumk-means} formulate balanced $k$-means clustering as a QUBO, achieving clustering quality on par with classical $k$-means and highlighting potential scalability benefits.
These proof-of-concept results illustrate that quantum annealing can effectively handle clustering objectives, optimizing over assignments of points to clusters in a manner that, in some cases, escapes local minima that trap classical heuristics.

Quantum clustering studies omitted explicit enforcement for spatial contiguity, as it requires a prohibitive number of qubits and scales poorly with the problem size.
They relied on adjacency graph structures that implicitly encoded connectivity.
Very recent work addresses spatial contiguity directly~\cite{amrQdata2025Paper} based on the discrete quadratic model (DQM)~\cite{dwave_leap_hybrid}.
However, none of the existing quantum-based clustering approaches have addressed seeding spatial regionalization, which is shown over the past decade to be the most scalable class of spatial regionalization algorithms.
\textbf{Our work is the first to explore seeding spatial regionalization on quantum computers.}

\vspace{-8pt}
\section{Quantum Spatial Regionalization}

Spatial regionalization is an NP-hard clustering problem, so all existing algorithms use heuristic-based search to find \textit{good approximate solutions} rather than optimal solutions.
Most existing seeding-based techniques work in two phases:
(1)~finding an initial solution, and
(2)~optimizing the initial solution to output a final solution.
Building on our previous work in~\cite{Alrashid023}, the first phase has four stages: \textit{seed selection}, \textit{region growing}, \textit{enclave assignment}, and \textit{constraint adjustments}.
The second phase is one stage: \textit{local optimization} that employs heuristic search techniques to improve the approximate solution quality, measured in regional heterogeneity.

We propose a hybrid approach that uses quantum annealing to improve the \textbf{\textit{seed selection}} and the \textbf{\textit{local optimization}} stages, while other stages are handled using efficient classical algorithms as in~\cite{Liu2022PRUC}.
These two stages have the highest potential to utilize the quantum annealing advantages to improve the runtime or solution quality.
The seed selection is purely quantum-encoded.
The local optimization is an iterative hybrid loop: the CPU prepares a set of potential region border areas, which the QPU then shuffles to refine the solution.
This ongoing work shows a preliminary, promising exploration that will be further strengthened in future explorations.
The rest of this section describes each quantum-encoded stage.

\vspace{-8pt}
\subsection{Quantum Seed Selection}

The seed selection stage takes $n$ spatial areas, i.e., polygons, as input and finds $p$ spatially scattered seed areas to build regions around.
Scatteredness ensures seeds remain well-separated so that subsequent regional growth is not squeezed together.
We measure scatteredness as the minimum pairwise distance between geometric centroids of seed areas, as in~\cite{Liu2022PRUC}, with the objective of maximizing this scatteredness.
Therefore, seed selection maps to a well-known problem called the max-min dispersion~\cite{yukiyoshi2024quantumdispersion}.
To address the max-min dispersion, we adopt a quantum-friendly formulation based on the Maximum Independent Set (MIS)~\cite{Ebadi2022}.
Given a set of 2D points \( V \), we first define a complete graph \( G = (V, E) \) where each edge \( e_{ij} \) is assigned a weight equal to the Euclidean distance \( d_{ij} \) between points \( i \) and \( j \).
The MIS is a set of vertices where no two vertices in the set are connected by an edge, and the set has the largest possible size among all such sets in the graph.
For a candidate minimum (dispersion) distance threshold \( d_m \), we construct a subgraph \( G' = (V, E') \) by removing all edges with \( d_{ij} > d_m \). Solving the MIS on \( G' \) yields a subset of points where all pairwise distances are at least \( d_m \).
If the MIS size is less than the required number of seeds \( p \), \( d_m \) is reduced; otherwise, it can be increased.
A binary search over feasible values of \( d_m \) identifies the largest achievable dispersion. We associate a binary variable $x_i$ with each spatial area, where $x_i=1$ if area $i$ is selected as a seed and $0$ otherwise. 
The primary objective is to maximize the number of seeds, which is equivalent to minimizing the term $-\sum_i x_i$. 
To enforce the MIS constraint, we introduce a penalty $\lambda_{\text{MIS}}$ for any pair of connected seeds being selected simultaneously. 
This results in the following objective function:
\[
X = - \sum_{i \in V} x_i + \lambda_{\text{MIS}}\sum_{(i,j) \in E'} x_i x_j
\label{eq:mis_bqm}
\]
Minimizing $X$ thus yields the largest possible set of seeds that are all separated by at least the distance threshold $d_m$.
This technique is encoded as a BQM model~\cite{dwave_leap_hybrid}.
For space limitations, our complete BQM formulation is described in~\cite{appndxQR}.

\subsection{Quantum Local Optimization}
\label{sec:local}

The local optimization stage takes an initial solution of $p$ spatially-contiguous regions as input, and outputs an optimized solution. 
The optimized solution still consists of $p$ spatially-contiguous regions but with lower \textit{regional heterogeneity}.
This stage is by far the most time-consuming stage for all seeding-based regionalization techniques, and it also determines the final objective function value, i.e., heterogeneity.
This makes it the best candidate to leverage the quantum advantage for faster runtime and better objective value.

We employ quantum annealing to model a “swap‑or‑move” local search technique that improves the initial solution by shuffling border movable areas between neighboring regions.
The goal is minimizing an objective function representing the overall \textbf{regional heterogeneity} while maintaining the \textbf{spatial contiguity} of each region. By finding articulation areas using the classic method~\cite{Liu2022PRUC}, we can find all movable areas in linear time. The heterogeneity of a region $R$ is defined as $H_R = \sum_{a_i,a_j \in R, i<j} |A(a_i) - A(a_j)|$, for the dissimilarity attribute $A$ of all pairs of areas $a_i, a_j \in R$.
The overall heterogeneity $H$ is the summation of heterogeneity of all the $p$ regions, i.e., $H = \sum_{k=1}^{p} H_{R_k}$, $\forall 1 \leq k \leq p$.
If an area $a_i$ moves from a donor region $R_d$ to a receiver region $R_r$, both $H_{R_d}$ and $H_{R_r}$ change, and hence $H$ changes.

We employ a Constraint Quadratic Model (CQM)~\cite{dwave_leap_hybrid} to optimize choosing area moves that improve the objective function $H$.
It is computationally expensive to aggregate the attribute $A$ over all area pairs directly on a QPU due to an architectural limitation of the qubits.
To overcome this, we adopt an estimation strategy that avoids full recalculation of $H$ by directly estimating the \textit{change} in heterogeneity ($\Delta_H$) resulting from a move.
The objective function is modeled using key statistical attributes of a region $R$, its total number of areas ($N_R$), mean attribute $A$ value ($M_R$), and attribute $A$ variance ($V_R$), along with the value of attribute $A_m$ of a movable border area $a_m \in R$.
These statistics are chosen for their computational efficiency, as they can be updated in linear or constant time on a classic computer.
The potential destinations for movable areas serve as the model’s variables. By assuming the attributes of large datasets are normally distributed, when a region adds or removes an area, this change is estimated using its statistical properties:
\[
\Delta \hat{H}_{\text{add}} = N_{R} \cdot \sqrt{\text{V}_R + (M_R - \text{A}_{m})^2}
\]
\[
\Delta \hat{H}_{\text{remove}} = -(N_R - 1) \cdot \sqrt{\text{V}_R + (M_R - \text{A}_{m})^2}
\]

$\Delta_H = \Delta \hat{H}_{\text{add}}+\Delta \hat{H}_{\text{remove}}$.
By estimating $\Delta_H$ for every potential move, we can leverage quantum optimization to simultaneously evaluate multiple relocations and select the combination that most significantly reduces total system heterogeneity.
To preserve spatial contiguity, we impose a critical constraint: each region may either donate or receive at most one area per iteration, but not both. This constraint is necessary because moving an area can alter the set of remaining movable areas and change the receivable areas of the donor region.

The hybrid quantum local optimization stage executes multiple iterations.
Each iteration computes the list of movable areas using the classical CPU.
Then, the list is dispatched to the QPU to explore multiple area moves using a CQM model to find a better solution.
Based on the separator theorem~\cite{Lipton77SeparatorTheorem}, the number of potential movable border areas ranges from $2\sqrt{np}$ to $2p\sqrt{n}$ in \textit{every single iteration}.
For space limitations, our CQM formulation is detailed in~\cite{appndxQR}.

\section{Empirical Results}

\setcounter{topnumber}{2}
\setcounter{bottomnumber}{2}
\setcounter{totalnumber}{4}
\renewcommand{\topfraction}{0.85}
\renewcommand{\bottomfraction}{0.85}
\renewcommand{\textfraction}{0.15}
\renewcommand{\floatpagefraction}{0.8}
\renewcommand{\textfraction}{0.1}
\setlength{\floatsep}{5pt plus 2pt minus 2pt}
\setlength{\textfloatsep}{5pt plus 2pt minus 2pt}
\setlength{\intextsep}{5pt plus 2pt minus 2pt}

\begin{figure}[t]
  \centering
  \begin{subfigure}{0.24\textwidth}
    \centering    \includegraphics[width=\linewidth]{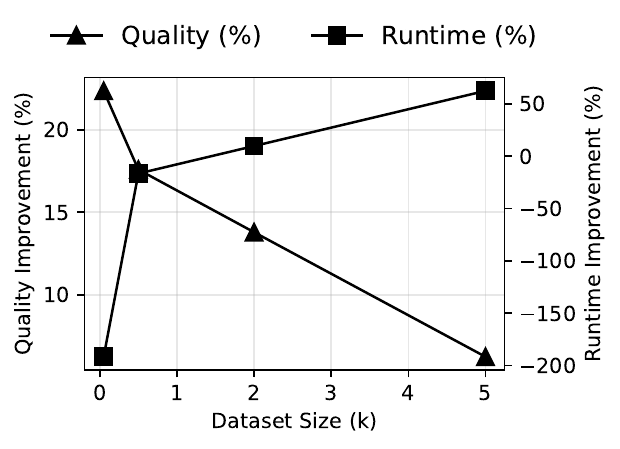}
    \caption{Quantum Improvements (\%)}
\label{fig:expr:local:improvements}
  \end{subfigure}
  \begin{subfigure}{0.22\textwidth}
    \centering
\includegraphics[width=\linewidth]{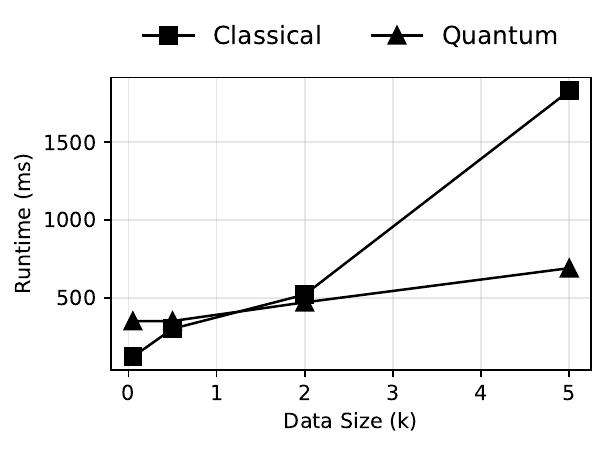}
    \caption{Absolute Runtime}
\label{fig:expr:local:runtim}
  \end{subfigure}
  \vspace{-14pt}
  \caption{Performance in local optimization vs. dataset size}  \label{fig:expr:local}
\end{figure}

\begin{figure}[t]
  \centering
  \begin{subfigure}{0.24\textwidth}
    \centering    \includegraphics[width=\linewidth]{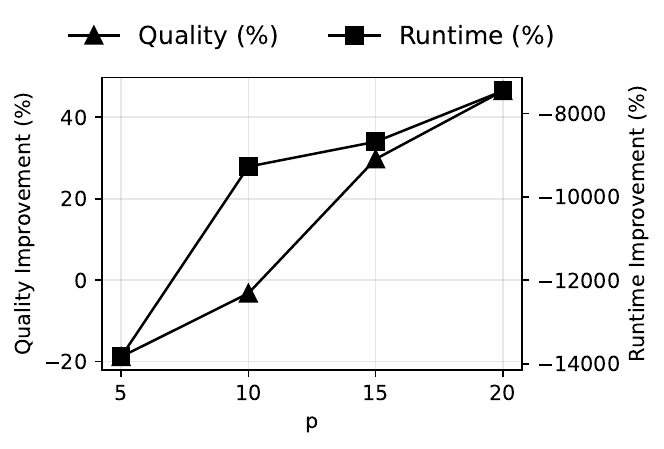}
    \caption{Quantum Improvements (\%)}
\label{fig:expr:seeds:improvements}
  \end{subfigure}
  \begin{subfigure}{0.22\textwidth}
    \centering
\includegraphics[width=\linewidth]{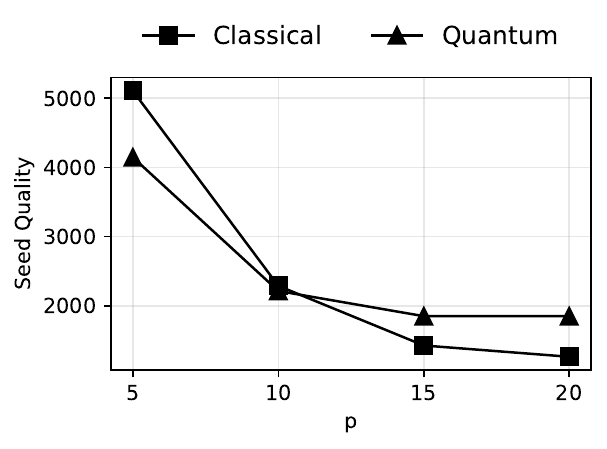}
    \caption{Absolute Quality}
\label{fig:expr:seeds:quality}
  \end{subfigure}
  \vspace{-14pt}
  \caption{Performance in seed selection vs. p}  \label{fig:expr:seeds}
\end{figure}

The proposed BQM and CQM models for \textit{seed selection} and \textit{location optimization}, respectively, are implemented using Python 3.12 and constructed using the D-Wave Ocean toolbox 'dimod' library (version 0.12.14).
We run our algorithms on an on-premise D-Wave Advantage annealing quantum computer (housed at the USC Quantum Computing Center) accessed via D-Wave's Leap cloud APIs.
The current QPU has Pegasus P16 topology and provides $\sim$5000 superconducting flux-qubits, each is 15-way connected.
We compare our quantum algorithms versus the classical counterparts in PRUC~\cite{Liu2022PRUC}, setting its user threshold to zero and using the same evaluation datasets.
The default value of dataset size is~50 and p is~10.
All results are collected after 10 iterations of local optimization.
The relative improvement of runtime is calculated as ($\frac{T_{classic}-T_{quantum}}{T_{classic}} \times 100$)\%, and similarly for quality improvements.

Figure~\ref{fig:expr:local} shows \textbf{\textit{initial results}} for the local optimization performance.
Figure~\ref{fig:expr:local:improvements} depicts the relative improvements, as percentages, of our quantum local optimization versus the classical counterpart in both runtime and quality, measured as the objective function value (overall regional heterogeneity $H$).
With increasing dataset size (DS), the runtime improvement increases while the quality improvement decreases.
The quality improvement starts at 22.4\% at DS=50 and continuously decreases to 6.3\% at DS=5k.
This suggests that quantum annealing keeps marginal improvements in solution quality with increasing data sizes, yet it will not degrade the quality.
On the contrary, the runtime improvement continuously increases from -191\% (a slowdown) at DS=50 to 62.3\% (a speedup) at DS=5k.
These results are confirmed in Figure~\ref{fig:expr:local:runtim}, which shows relatively stable quantum runtime as data size increases.
This shows a promising quantum advantage in terms of runtime speedup with larger datasets.
It worth noting that a dataset of all US counties has 3.2k polygons, so 5k datasets are used in many practical applications.

Figure~\ref{fig:expr:seeds} shows \textbf{\textit{initial results}} for the seed selection performance.
Figure~\ref{fig:expr:seeds:improvements} depicts the relative improvements, as percentages, showing better runtime and quality improvements with increasing $p$ value.
However, the quantum runtime ranges from 300- 500 milliseconds compared to 3-7~milliseconds for the classical algorithm for this small dataset size, so the increasing improvement is not significant and must be tested on larger data sizes that our preliminary model does not currently support.
On the contrary, the quality improvement, measured as the min seed distance, is significantly improved to reach 46.5\%.
This is confirmed by the absolute quality values in Figure~\ref{fig:expr:seeds:quality} as larger $p$ makes it harder for classical heuristics to optimize for many seeds, while the quantum annealer can still explore a larger search space.
This suggests a potential quantum advantage in improving the seed selection quality.

\vspace{-8pt}
\section{Conclusion and Future Explorations}

This paper has presented the first hybrid quantum approach to address the spatial regionalization problem.
The approach decomposes the regionalization process into stages, and exploits the quantum computing power for the two most promising stages, seed selection and local optimization.
Two algorithms have been proposed and realized using D-Wave's BQM and CQM hybrid solvers.
The presented initial results suggest a potential quantum advantage to improve the objective function quality of seed selection and the runtime efficiency of the local optimization.
More explorations are ongoing to improve the models in terms of scalability, more effective heuristics, and support richer variants of spatial regionalization problems.
Future extensions can use our quantum model to enforce spatial contiguity~\cite{amrQdata2025Paper} towards a quantum \textit{region growing} stage.

\vspace{-8pt}
{\footnotesize
\bibliographystyle{abbrv}
{\footnotesize\bibliography{sample-base,amr,references_sabek_qdata24}}

\begin{thebibliography}{10}

\bibitem{appndxQR}
{Appendix}.
\newblock \url{http://cs.ucr.edu/~amr/AppendixQR.pdf}, 2025.

\bibitem{Alrashid2022SMP}
H.~Alrashid, Y.~Liu, and A.~Magdy.
\newblock {SMP: Scalable Max-P Regionalization}.
\newblock In {\em SIGSPATIAL}, 2022.

\bibitem{AlrashidLM23}
H.~Alrashid, Y.~Liu, and A.~Magdy.
\newblock {PAGE:} parallel scalable regionalization framework.
\newblock {\em {ACM} Transactions on Spatial Algorithms and Systems}, 9(3):1--26, 2023.

\bibitem{Alrashid023}
H.~Alrashid and A.~Magdy.
\newblock {A Scalable Unified System for Seeding Regionalization Queries}.
\newblock In {\em Proceedings of the 18th International Symposium on Spatial and Temporal Data, {SSTD}}, pages 96--105, 2023.

\bibitem{Quantumk-means}
D.~Arthur and P.~Date.
\newblock Balanced \emph{k}-means clustering on an adiabatic quantum computer.
\newblock {\em Quantum Information Processing}, 20(9):294, 2021.

\bibitem{amrQdata2025Paper}
Y.~Chang, A.~Magdy, and F.~M. Spedalieri.
\newblock {Quantum Modeling of Spatial Contiguity Constraints}, 2025.
\newblock https://arxiv.org/abs/2505.12608. To appear at the ACM SIGMOD Workshop on Quantum Computing and Quantum-Inspired Technology for Data-Intensive Systems and Applications (Q-Data 2025).

\bibitem{web:dwavesuccessstories}
D-Wave.
\newblock {Real-World Quantum Applications at Business Scale}.
\newblock \url{https://www.dwavequantum.com/learn/customer-success-stories/}, 2025.

\bibitem{dwave_leap_hybrid}
{D-Wave Systems Inc.}
\newblock {\em Leap Service’s Hybrid Solvers}.
\newblock D‑Wave Systems Inc., 2025.
\newblock Accessed 2025‑06‑12.

\bibitem{Ebadi2022}
S.~Ebadi, T.~T. Wang, H.~Levine, A.~Keesling, G.~Semeghini, K.~Pichler, H.~Bernien, et~al.
\newblock Quantum optimization of maximum independent set using rydberg atom arrays.
\newblock {\em Science}, 376(6598):1209--1215, 2022.

\bibitem{web:geoda_spatial_clustering}
{GeoDa Center for Geospatial Analysis and Computation}.
\newblock {Spatial Clustering — pygeoda 1.0.0 documentation}.
\newblock \url{https://geodacenter.github.io/pygeoda/spatial_clustering.html}, 2025.
\newblock Accessed: 2025-06-12.

\bibitem{Glover2022QUBO}
F.~Glover, G.~Kochenberger, R.~Hennig, and Y.~Du.
\newblock Quantum bridge analytics i: a tutorial on formulating and using qubo models.
\newblock {\em Annals of Operations Research}, 314(1):141--183, 2022.

\bibitem{kadowaki1998}
T.~Kadowaki and H.~Nishimori.
\newblock Quantum annealing in the transverse ising model.
\newblock {\em Phys. Rev. E}, 58:5355--5363, Nov 1998.

\bibitem{Kang2022EMP}
Y.~Kang and A.~Magdy.
\newblock {EMP: Max-P Regionalization with Enriched Constraints}.
\newblock In {\em 2022 IEEE 38th International Conference on Data Engineering (ICDE)}, 2022.

\bibitem{King2023Quantum}
A.~D. King et~al.
\newblock Quantum critical dynamics in a 5,000-qubit programmable spin glass.
\newblock {\em Nature}, 617(7959):61--66, 2023.

\bibitem{Koshka2020DWave}
Y.~Koshka and M.~A. Novotny.
\newblock {Comparison of D-Wave Quantum Annealing and Classical Simulated Annealing for Local Minima Determination}.
\newblock In {\em Proceedings of the 2020 IEEE International Conference on Rebooting Computing (ICRC)}, 2020.

\bibitem{Lipton77SeparatorTheorem}
R.~J. Lipton and R.~E. Tarjan.
\newblock {A Separator Theorem for Planar Graphs}.
\newblock {\em SIAM Journal on Applied Mathematics}, 36(2):177--189, 1979.

\bibitem{Liu2022PRUC}
Y.~Liu, A.~R. Mahmood, A.~Magdy, and S.~Rey.
\newblock {PRUC: P-regions with User-Defined Constraint}.
\newblock {\em Proc. VLDB Endow.}, 15(3), 2021.

\bibitem{QACommunity2020Negre}
C.~Negre, H.~Ushijima-Mwesigwa, and S.~Mniszewski.
\newblock Detecting multiple communities using quantum annealing on the d-wave system.
\newblock {\em PLOS ONE}, 15, 2020.

\bibitem{QAGraph2017Ushijima}
H.~Ushijima-Mwesigwa, C.~F.~A. Negre, and S.~M. Mniszewski.
\newblock Graph partitioning using quantum annealing on the d-wave system.
\newblock In {\em Proceedings of the Second International Workshop on Post Moores Era Supercomputing}, PMES'17, 2017.

\bibitem{yukiyoshi2024quantumdispersion}
K.~Yukiyoshi, T.~Mikuriya, H.~S. Rou, G.~T.~F. de~Abreu, and N.~Ishikawa.
\newblock Quantum speedup of the dispersion and codebook design problems.
\newblock {\em IEEE Transactions on Quantum Engineering}, 2024.

\end{thebibliography}
}
\appendix
\section*{Appendix}
\section{BQM in Seed Selection}
\label{sec:bqm}

The seed selection stage is modeled as a max-min dispersion problem, which we solve by repeatedly finding the Maximum Independent Set (MIS) in a graph. We solve this by employing a quantum-assisted binary search to find the largest possible minimum distance ($d_m$) that allows for a valid selection of $p$ seeds.


\vspace{4pt}
\textbf{Variables.} Let $V = \{a_1, a_2, \ldots, a_n\}$ be the set of spatial areas. We define a binary variable for each area:
$$
x_i = \begin{cases}
1 & \text{if area } a_i \text{ is selected as a seed} \\
0 & \text{otherwise}
\end{cases}
$$

\textbf{Objective Function.} For a given minimum distance threshold $d_m$, we construct a graph $G' = (V, E')$ where edge $(i,j) \in E'$ if $d'_{ij} \leq d_m$. The MIS objective maximizes the number of selected areas:
$$
\text{Maximize} \quad \sum_{i=1}^{n} x_i
$$
By keep finding MIS qualified for a minimum distance, we can find a more precise range of $d_m$ by iteration.

\vspace{4pt}
\textbf{Constraints:}
To ensure we select exact $p$ seeds from dataset and try to make them scattered, we enforce the following constraints:
\paragraph{MIS Constraint} No two adjacent areas in $G'$ can both be selected as seeds. For each edge $(i,j) \in E'$:
$$
x_i \cdot x_j = 0
$$
This is enforced by adding a quadratic penalty term $\lambda_{\text{MIS}} \cdot x_i \cdot x_j$ to the objective, where $\lambda_{\text{MIS}}$ is a sufficiently large penalty coefficient. The overall objective function with constraints penalties is:
\[
X = - \sum_{i \in V} x_i + \lambda_{\text{MIS}} \sum_{(i,j) \in E'} x_i x_j
\label{eq:mis_bqm:appndx}
\]

The requirement of selecting exactly $p$ seeds is handled by a classical binary search that repeatedly calls this BQM. After the BQM returns a maximum independent set of size $k$, if $k \ge p$, the distance $d_m$ is feasible; we increase it to seek better dispersion. If $k < p$, the distance $d_m$ is too high; we must decrease it. This process converges on the maximum $d_m$ that supports a solution of at least $p$ seeds.
If it converges to more than $p$, the final output $p$ seeds are selected at random.

\section{CQM in Location Optimization}
\label{sec:cqm}

To run the local optimization phase on a D-Wave system, we designed a Constrained Quadratic Model (CQM).  This model is defined by three key components: binary variables representing all areas' potential moves calculated by CPU~\cite{Liu2022PRUC}, an objective function to estimate the change in heterogeneity, and constraints to preserve spatial contiguity.

\vspace{4pt}
\textbf{Variables.} Let $S$ be the set of all possible moves, where a move is a triplet $(i, d, r)$ representing the transfer of area $a_i$ from a donor region $R_d$ to a receiver region $R_r$. We define a binary variable for each possible move:
$$
x_{i,d,r} = \begin{cases}
1 & \text{if area } a_i \text{ is moved from } R_d \text{ to } R_r \\
0 & \text{otherwise}
\end{cases}
$$

Based on the separator theorem~\cite{Lipton77SeparatorTheorem}, the number of potential movable border areas ranges from $2\sqrt{np}$ to $2p\sqrt{n}$ in \textit{every single iteration}.
This is a major number for large values of $n$, which is typically associated with larger $p$ values as well.

\vspace{4pt}
\textbf{Objective Function.}
The objective is to minimize the total estimated change in heterogeneity across all selected moves. This is formulated as:
$$
\text{Minimize} \quad \sum_{(i,d,r) \in S} \left( \Delta \hat{H}_{\text{add}} + \Delta \hat{H}_{\text{remove}} \right) \cdot x_{i,d,r}
$$
The terms $\Delta \hat{H}_{\text{add}}$ and $\Delta \hat{H}_{\text{remove}}$ are calculated for each potential move based on the statistical properties of the involved regions, as defined in Section~\ref{sec:local}.

\vspace{4pt}
\textbf{Constraints.} To ensure the solution is connectivity-stable during areas exchanging, we impose the following constraints:

\paragraph{1. One Move per Area} Each movable area $a_i$ can be transferred at most once. This is enforced for every area $a_i$ in the set of movable areas. Given that for a specific area $a_i$, its donor region $R_d$ is fixed, we sum over all possible receiver regions $R_r$:
$$
\forall a_i \in \text{MovableAreas}, \quad \sum_{r \mid (i,d,r) \in S} x_{i,d,r} \le 1
$$

\paragraph{2. Region Stability} Each region $R_k$ can either act as a donor for one move or as a receiver for one move, but not both. This is enforced for every region $k$ from $1$ to $p$:
$$
\forall k \in \{1, \ldots, p\}, \quad {\sum_{(i,k,r) \in S} x_{i,k,r}} + {\sum_{(i,d,k) \in S} x_{i,d,k}} \le 1
$$
This constraint ensures that all selected moves are mutually compatible within a single iteration, preventing cascading changes that could invalidate the premises of other concurrent moves.

\end{document}